\documentclass[aps,prl,twocolumn,showpacs,showkeys,footinbib,amsmath,amssymb,superscriptaddress]{revtex4-1}

\usepackage{amsmath}
\usepackage{amssymb}
\usepackage{graphicx}
\usepackage{bm}
\usepackage{color}
\usepackage{relsize}
\usepackage{braket}
\usepackage{bbold}
\usepackage{txfonts}
\usepackage{epstopdf}
\usepackage{hyperref}

\begin{document}

\title{Quantum Spin-Valley Hall Kink States: From Concept to Materials Design}

\author{Tong Zhou}
\email{tzhou8@buffalo.edu}
\affiliation{Department of Physics, University at Buffalo, State University of New York, Buffalo, New York 14260, USA}
\author{Shuguang Cheng}
\affiliation{Department of Physics, Northwest University, Xi'an 710069, China}
\author{Michael Schleenvoigt}
\affiliation{Peter Gr{\"u}nberg Institute 9, Forschungszentrum J{\"u}lich $\&$ JARA J{\"u}lich-Aachen Research Alliance, 52425 J{\"u}lich, Germany}
\author{Peter Sch{\"u}ffelgen}
\affiliation{Peter Gr{\"u}nberg Institute 9, Forschungszentrum J{\"u}lich $\&$ JARA J{\"u}lich-Aachen Research Alliance, 52425 J{\"u}lich, Germany}
\author{Hua Jiang}
\affiliation{School of Physical Science and Technology, Soochow University, Suzhou 215006, China}
\author{Zhongqin Yang}
\affiliation{State Key Laboratory of Surface Physics and Key Laboratory of Computational Physical Sciences (MOE) and Department of
Physics and Collaborative Innovation Center of Advanced Microstructures, Fudan University, Shanghai 200433, China}
\author{Igor \v{Z}uti\'c}
\email{zigor@buffalo.edu}
\address{Department of Physics, University at Buffalo, State University of New York, Buffalo, New York 14260, USA}
\date{\today}
\begin{abstract}
We propose a general and tunable platform to realize high-density arrays of quantum spin-valley Hall kink (QSVHK) states with spin-valley-momentum locking based on a two-dimensional hexagonal topological insulator. Through the analysis of Berry curvature and topological charge, the QSVHK states are found to be topologically protected by the valley-inversion and time-reversal symmetries. Remarkably, the conductance of QSVHK states remains quantized against both nonmagnetic short/long-range and magnetic long-range disorder, verified by the Green function calculations. Based on first-principles results and our fabricated samples, we show that QSVHK states, protected with a gap up to 287$\,$meV, can be realized in bismuthene by alloy engineering, surface functionalization, or electric field, supporting non-volatile applications of spin-valley filters, valves, and waveguides even at room temperature. 
\end{abstract}
\maketitle

% a, k definition
Two-dimensional (2D) hexagonal lattices offer a versatile platform to manipulate charge, spin, and valley degrees freedom and implement different topological states. While pioneering predictions for quantum anomalous and quantum spin Hall (QSH) effect~\cite{Haldane1988:PRL,Kane2005:PRL} were guided by graphene-like systems, graphene poses inherent difficulties with its weak spin-orbit coupling (SOC) and a gap of only 
$\Delta \sim 40 \,\mu$eV~\cite{Sichau2019:PRL}. 
The quest for different 2D hexagonal monolayers (MLs) with a stronger SOC on one hand reveals, as in transition metal dichalcogenides (TMDs), an improved control of valley-dependent phenomena~\cite{Xu2014:NP}, emulating extensive research in spintronics~\cite{Zutic2004:RMP}, while on the other hand, as in ML Bi on SiC substrate, topological states remain even above the room temperature, with a huge topological gap $\sim$ 0.8 eV~\cite{Reis2017:S}. However, examples where valley degrees of freedom support robust topological states are scarce.

In 2D materials with broken inversion symmetry, such as gapped 
graphene~\cite{Xiao2007:PRL,Qiao2011:PRL,McCann2013:RPP,Gorbachev2014:S,Sui2015:NP,Shimazaki2015:NP} and TMDs~\cite{Xiao2012:PRL,Mak2014:S}, the opposite sign of the momentum-space Berry curvature $\mathbf{\Omega}(\mathbf{k})$ in different valleys is responsible for a valley Hall effect, where the carriers in different valleys turn into opposite directions transverse to an in-plane electric field~\cite{Xiao2007:PRL,Mak2014:S}. A striking example of such a sign reversal in $\mathbf{\Omega}(\mathbf{k})$ along an internal boundary of a film is realized in quantum valley Hall kink (QVHK) states ~\cite{Martin2008:PRL,Jung2011:PRB,Qiao2011:NL,Vaezi2013:PRX,Zhang2013:PNAS,Ju2015:N,Li2016:NN,Yin2016:NC,
Li2018:S,Chen2020:NC,Semenoff2008:PRL,Kim2014:PRB,Cheng2018:PRL,Hu2018:PRL,Wang2021:Nanotechnology}.  
The resulting topological defect supports counterpropagating 1D chiral electrons, topologically protected by the valley-inversion symmetry~\cite{Martin2008:PRL,Jung2011:PRB,Qiao2011:NL,Vaezi2013:PRX,Zhang2013:PNAS}. The underlying mechanism for the formation of zero-energy states, expected from the index theorem~\cite{Shen:2012,Junker:1996}, shares similarities with many other systems in condensed matter and 
particle physics~\cite{Jackiw1976:PRD,Su1979:PRL,Adagideli1999:PRL,Sengupta2001:PRB,Lee2007:PRL,Nishida2010:PRB,Yasuda2017:Science,Sedlmayr2020:PRB}.
While the proposals for QVHK mainly focus on bilayer graphene (BLG) systems~\cite{Ju2015:N,Li2016:NN,Yin2016:NC,Li2018:S,Chen2020:NC},
the required sign reversal in $\mathbf{\Omega}(\mathbf{k})$ realized by either the random local stacking faults~\cite{Ju2015:N,Yin2016:NC}, or a dual-split-gate structure~\cite{Li2016:NN,Li2018:S,Chen2020:NC}, is challenging to implement to achieve high-density channels. With the required applied electric field, the volatility of QVHK limits their envisioned use in valleytronics. A small gap of BLG $\sim$ 20 meV~\cite{Li2016:NN} excludes high-temperature applications, and QVHK states were limited to 5 K~\cite{Ju2015:N,Li2016:NN,Yin2016:NC,Li2018:S,Chen2020:NC}. 
Crucially, disorder easily induces intervalley scattering, preventing the expected ballistic transport in QVHK states~\cite{Ju2015:N,Li2016:NN,Wang2021:Nanotechnology}.

\begin{figure}[t]
\includegraphics*[width=0.46\textwidth]{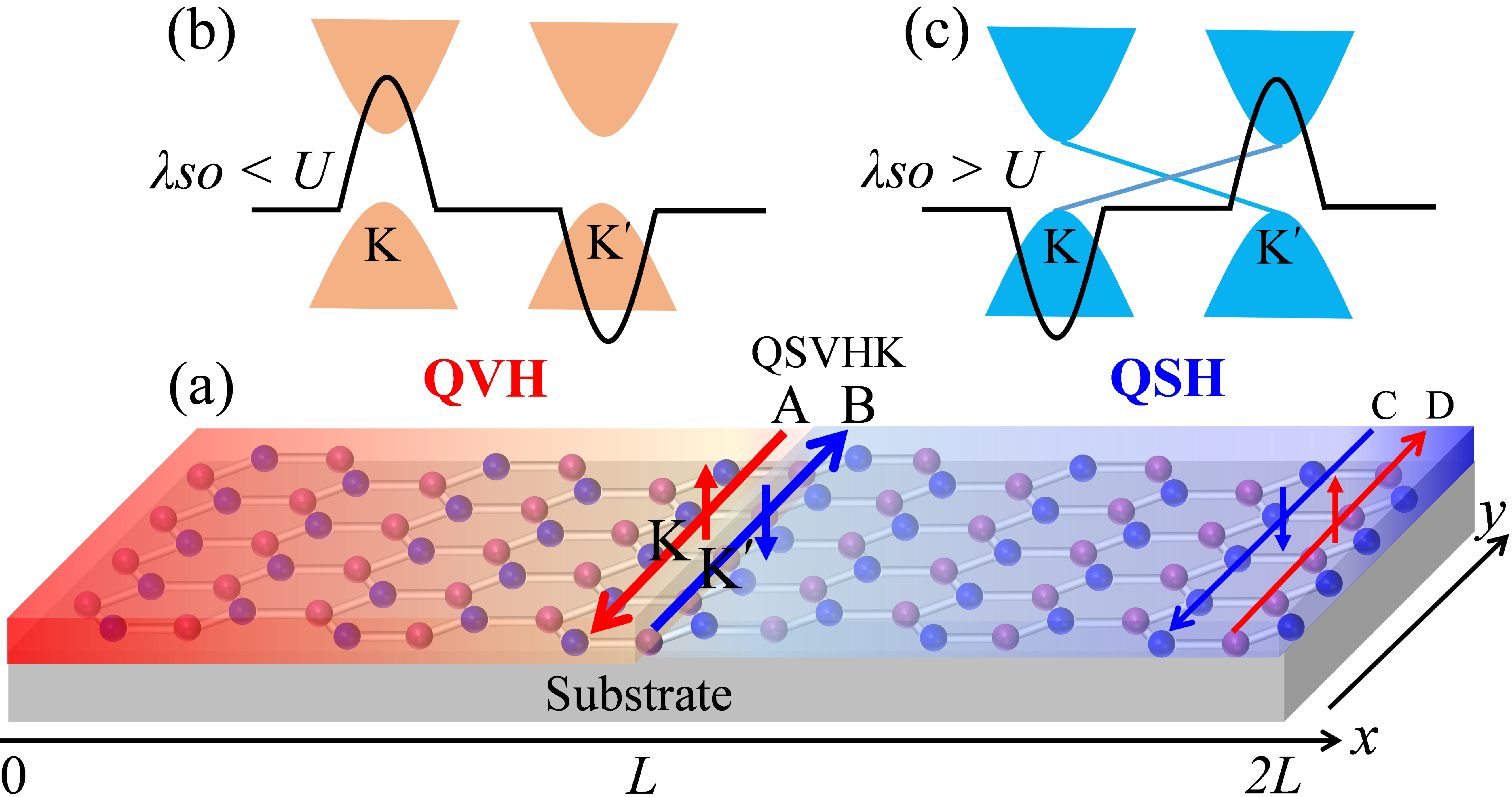}
\caption{(a) Schematic of the QSVHK states ($A$, $B$) at the valleys $K$ and $K'$, and QSH edge states ($C$, $D$) in a junction formed by QVH and QSH insulators. 2$L$ is the junction length. The red (blue) arrow denotes the spin-up (down) channel. (b) and (c) the schematic of the bands and Berry curvatures (black lines) for QVH and QSH insulators, distinguished
by the  relative strength of SOC, $\lambda_{SO}$, and staggered potential, $U$.}
\label{fig:F1}
\end{figure}
 
\begin{figure*}[t]
\centering
\includegraphics*[width=0.74\textwidth]{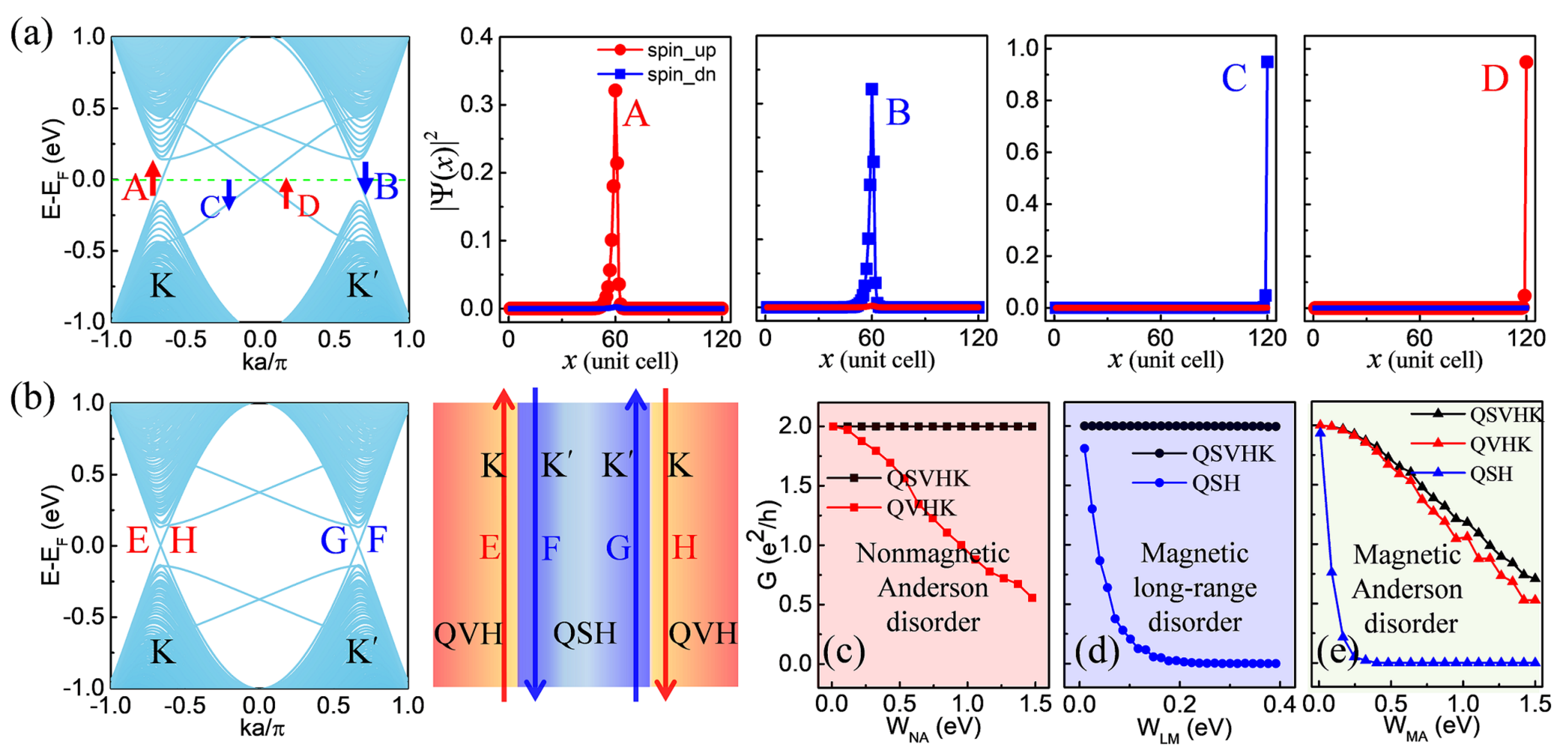}
\caption{(a)
Bands and wave function distributions (${|\Psi(x)|}^{2}$) for topological states $A$-$D$ in the QSH-QVH junction with $L = 60$ unit cells. (b) Bands and schematic of QVH-QSH-QVH junction with pure QSVHK states $E$-$H$. (c) Junction conductance,$G$, versus nonmagnetic Anderson disorder strength $W_{NA}$ at the Fermi level, for the QSVHK and QVHK. (d) Same as (c) but for the QSVHK and QSH versus magnetic long-range disorder strength $W_{LM}$. (e) Same as (c) but versus magnetic Anderson disorder strength $W_{MA}$.  The parameters $U$, $\lambda_{SO}$ for QVH and QSH regions are taken from the BiAs/SiC and Bi/SiC, respectively. The hopping parameter $t_{1,2}=1\,$eV.}
\label{fig:F2}
\end{figure*} 
 
Motivated by these challenges, we propose a robust platform to realize high-density arrays of spin-polarized QVHK states at room temperature based on a 2D hexagonal topological insulator, where the QVHK states are simultaneously the QSH edge states, forming along the QSH-QVH interface as shown in Fig.~1. The QSH is described by a topological invariant $Z_2=1$~\cite{Kane2005:PRL}, while the QVH is characterized by a quantized valley Chern number $C_V = 1$ and $Z_2=0$~\cite{Zhang2013:PNAS,Qiao2011:PRL}. Across their interface, both $Z_2$ and $C_V$ change the value, thus the QSH and QVHK states simultaneously emerge along the interface, giving largely unexplored topological kink states, we term quantum spin-valley Hall kink (QSVHK) states. Unlike the previous studies of the interplay between topological states~\cite{Martin2008:PRL,Jung2011:PRB,Qiao2011:NL,Vaezi2013:PRX,Zhang2013:PNAS,Ju2015:N,Li2016:NN,Yin2016:NC,
Li2018:S,Chen2020:NC,Semenoff2008:PRL,Kim2014:PRB,Cheng2018:PRL,Hu2018:PRL,Wang2021:Nanotechnology,Pan2014:PRL,Zhou2015:NL,Zhou2016:PRB,Zhou2017:PRL,
Tahir2013:APL,Huang2015:PRB,Banerjee2021:PRB,
Nishida2010:PRB,Nishida2010:PRB,Yasuda2017:Science,Lee2007:PRL,Sedlmayr2020:PRB}, our proposed QSVHK shows a peculiar marriage between QSH and QVH. The QSH becomes robust against the magnetic long-range disorder due to the valley-inversion symmetry protection of the QVH~\cite{Martin2008:PRL,Jung2011:PRB,Qiao2011:NL,Vaezi2013:PRX,Zhang2013:PNAS}, while the QVH can be robust against the nonmagnetic short-range disorder because of the time-reversal symmetry protection of the QSH~\cite{Kane2005:PRL,Qi2011:RMP}. Thus, in contrast to the trivial spin-valley Hall effects~\cite{Islam2016:Carbon,Dyrdal2017:2DM,Zhou2018:npjQM}, the topological QSVHK states are robust against both nonmagnetic short/long-range and magnetic long-range disorder, giving robust ballistic spin-valley-momentum locking transport. The QSH-QVH junction can be implemented by inducing a staggered potential, $U$, in a part of the 2D topological insulators. When $U$ is smaller (larger) than the strength of the intrinsic SOC, $\lambda_{SO}$, the QSH (QVH) is obtained. Our first-principles calculations reveal that $U$ can be induced by alloy engineering or surface decoration and easily controlled by the electric filed. 

We first present our idea through the analysis of a tight-binding model based on $p_x$ and $p_y$ orbitals, which is widely used to describe the physics of the  hexagonal MLs including arsenene~\cite{Zhang2018:CSR,Gui2019:JMCA}, antimonene~\cite{Ji2016:NC,Zhou2015:NL,Zhou2016:PRB,Xu2020:PRL}, bismuthene~\cite{Reis2017:S,Song2014:NPGAM,Zhou2018:npjQM}, and binary element group-V MLs~\cite{Liu2019:MH,Li2017:SR}
\begin{equation}
\begin{aligned} 
H=\lambda_{SO} \sum_{i} c_{i}^{+} \sigma_{z} \otimes s_{z} c_{i}+\sum_{i} U_{i} c_{i}^{+} \sigma_{0} \otimes s_{0} c_{i}\\+
\left(\sum_{i} \sum_{j=1,2,3} c_{i}^{+} T_{\delta_{j}} c_{i+\delta_{j}}+H . C .\right).
\end{aligned} 
\label{eq:BdG}
\end{equation} 
Here, $c_i$ represents the annihilation operator on site $i$. $\sigma$ and $s$ indicate the Pauli matrices acting on orbital and spin spaces. The first term describes the intrinsic SOC and the second term gives the staggered potential with $U_{i}=U(-U)$ for the  $A(B)$ sublattice. The hopping term 

\begin{equation} 
T_{\delta_{j}}=\left[\begin{array}{cc}
t_{1} & z^{(3-j)} t_{2} \\
z^{j} t_{2} & t_{1}
\end{array}\right] \otimes s_{0},
\end{equation}
describes the nearest hopping from site $i$ to $i + \delta_j$, where $z$=exp$(2i\pi/3)$ and $t_{1/2}$ is the hopping coefficient. In the absence of the first two terms in Eq.~(1), the gapless Dirac points exist at the two valleys~\cite{Song2014:NPGAM,Zhou2018:npjQM}. The staggered potential and intrinsic SOC open a gap of $2\left|\lambda_{SO}-\mathrm{U}\right|$ at the Dirac points and their competition determines the topology of the system. When $\lambda_{SO} < U$, the system is in QVH with $Z_2=0$ and opposite $\mathbf{\Omega}(\mathbf{k})\neq0$ at the two valleys [Fig.~1(b)]. When $\lambda_{SO} > U$, the system is in QSH with $Z_2=1$ and for $U >0$ the $\mathrm{sgn}[\mathbf{\Omega}(\mathbf{k})]$ is reversed as compared to the QVH [Fig.~1(c)]. We consider a planar junction formed by the QVH and QSH [Fig.~1(a)], where QSVHK emerges along their interface, since both $Z_2$ and $\mathbf{\Omega}(\mathbf{k})$ change the sign.

To identify such QSVHK states, we calculate the spectrum of the QSH-QVH junction along the zigzag direction, where the valley degree can be preserved~\cite{Ren2016:RPP,Avalos-Ovando2019:PRB,Wang2021:Nanotechnology}. As shown in Fig.~2(a), there are four non-degenerate gapless states, $A-D$, in the bulk band gap. The helical $C$ and $D$ states are the common QSH states localized at the outer edge of the QSH region, verified by their wave functions in Fig.~2(a). The $A$ ($B$) state at the $K$ ($K^{\prime}$) valley shows the QSVHK localized at the inner interface [Fig.~2(a)]. Unlike the QVHK in BLG, the QSVHK  is fully spin-polarized. Specifically, the kink state $A$ ($B$) at $K$ ($K^{\prime}$) valley has spin-up (down) channel. Such spin-valley-momentum locking supports a perfect spin-valley filter.

To better understand the emergence of the QSVHK, we focus on the low-energy physics of Eq.(1). We expand the Hamiltonian around the valleys and obtain a continuum model
\begin{equation}
H=\hbar v_{F}\left(k_{x} \sigma_{x}+\tau_{z} k_{y} \sigma_{y}\right)+\lambda_{SO} s_{z} \tau_{z} \sigma_{z}+U \sigma_{z},
\label{eq:cont}
\end{equation}
where $\nu_F$ is the Fermi velocity, $\sigma$, $s$, and $\tau$ are Pauli matrix for orbital, spin, and valley, respectively. From the index theorem~\cite{Shen:2012}, the number of the kink channels is related to the change of the bulk topological charges across the interface~\cite{Martin2008:PRL,Zhang2013:PNAS,Pan2015:PRB}. The spin- and valley-projected topological charge $C_{\tau_{z}}^{s_{z}}$ can be calculated by integrating the spin-dependent $\mathbf{\Omega}(\mathbf{k})$ of the valence bands around each valley ~\cite{Martin2008:PRL,Zhang2013:PNAS,Pan2015:PRB}. From the continuum model in Eq.~(3), we obtain   
\begin{equation}
C_{\tau_{z}}^{s_{z}}=\frac{\tau_{z}}{2} \operatorname{sgn}\left(U-\tau_{z} s_{z} \lambda_{SO}\right).
\end{equation} 
In the QVH region, we get $\left(C_{K}^{\uparrow}, C_{K}^{\downarrow}, C_{K^{\prime}}^{\uparrow}, C_{K^{\prime}}^{\downarrow}\right)$ = (0.5, 0.5, -0.5, -0.5), while in the QSH region, $\left(C_{K}^{\uparrow}, C_{K}^{\downarrow}, C_{K^{\prime}}^{\uparrow}, C_{K^{\prime}}^{\downarrow}\right)$ = (-0.5, 0.5, -0.5, 0.5). The number of the kink modes per spin/valley $\left(\nu_{K}^{\uparrow}, \nu_{K}^{\downarrow}, \nu_{K^{\prime}}^{\uparrow}, \nu_{K^{\prime}}^{\downarrow}\right)$ is an integer evaluated from the difference between the topological charges in two regions~\cite{Martin2008:PRL,Zhang2013:PNAS}, i.e., $\left(\nu_{K}^{\uparrow}, \nu_{K}^{\downarrow}, \nu_{K^{\prime}}^{\uparrow}, \nu_{K^{\prime}}^{\downarrow}\right)$ = (1, 0, 0, -1). It is clear the spin-up (down) topological charge has an integer change at the $K$ ($K^{\prime}$) valley, giving the spin-valley polarized QSVHK. This topological charge analysis is consistent with our discussion about the energy spectrum in Fig.~2(a). In QSH-QVH junctions, there are still QSH states along the outer edge. To eliminate them and realize a pure QSVHK transport, we propose in Fig.~2(b) a QVH-QSH-QVH junction, where the two pairs of QSVHK are verified by the calculated bands. Multiple channels can be expected with more QSH-QVH boundaries, where the width of each region should be large enough to avoid the interplay between adjacent QSVHK states.

For valley-related transport, the influence of the short- (long)-range disorder is usually significantly different since the former (latter) induces (excludes) intervalley scattering~\cite{Sarma2011:RMP}.  The former (latter) is characterized by the smaller (larger) disorder correlation length, $\lambda$, compared to the lattice spacing, $a$~\cite{Sarma2011:RMP}. For example, the  QVHK is only robust against the long-range disorder~\cite{Martin2008:PRL,Jung2011:PRB,Qiao2011:NL,Vaezi2013:PRX,Zhang2013:PNAS,
Ju2015:N,Li2016:NN,Yin2016:NC,Li2018:S,Chen2020:NC}. 
To explore the robustness of the QSVHK against the disorder, we calculate the junction conductance, $G$, using the Landauer-B{\"u}ttiker formula~\cite{Datta:1997} and the Green-function method~\cite{Sancho1984:JPF,Jiang2009:PRB,Qiao2016:PRL,Cheng2018:PRB} in the presence of nonmagnetic Anderson disorder ($\lambda\rightarrow$0)~\cite{Anderson1958:PR,Jiang2009:PRB} in the energy range (-$W_{NA}$/2, $W_{NA}$/2), magnetic Anderson disorder ~\cite{Montgomery1970:PRL,Qiao2016:PRL} (-$W_{MA}$/2, $W_{MA}$/2), and  magnetic long-range ($\lambda = 7a$) disorder ~\cite{Rycerz2007:EPL,Cheng2016:NJP} (-$W_{LM}$/2, $W_{LM}$/2), where $W_{NA}$, $W_{MA}$, and $W_{LM}$ measure their respective strengths. For comparison, we calculate $G(W_{NA})$, $G(W_{LM})$ and $G(W_{MA})$ in QVHK and QSH states, shown in Figs.~2(c)-(e). See calculation details and the crossover between the short- and long-range disorder in Supplemental Material (SM)~\cite{SMkink}. For the QVHK with valley-momentum locking, its $G$ decreases with $W_{NA}$ increases, consistent with previous studies~\cite{Ju2015:N,Li2016:NN}, because the Anderson disorder breaks the valley-inversion symmetry and leads to the intervalley scattering. For QSH with spin-momentum locking, its $G$ decreases with $W_{LM}$ ($W_{MA}$), because the time-reversal symmetry is broken by the magnetic disorder, in agreement with the experiments~\cite{Konig2007:S,Culcer2020:2DM}. In contrast, for QSVHK with spin-valley-momentum locking protected by both valley-inversion and time-reversal symmetries, its $G$ remains quantized against both nonmagnetic Anderson disorder and magnetic long-range disorder [Figs. 2(c)-(d)]. The backscattering in QSVHK can only be induced by simultaneously breaking the valley-inversion and time-reversal symmetries, for example by magnetic Anderson disorder [Fig. 2(e)]. However, with $W_{MA}$, the $G$ of QSVHK is still higher than that of the QSH and QVH, since simultaneously scattering spin and valley is harder than scattering each of them. 

\begin{figure}[t]
\centering
\includegraphics*[width=0.47\textwidth]{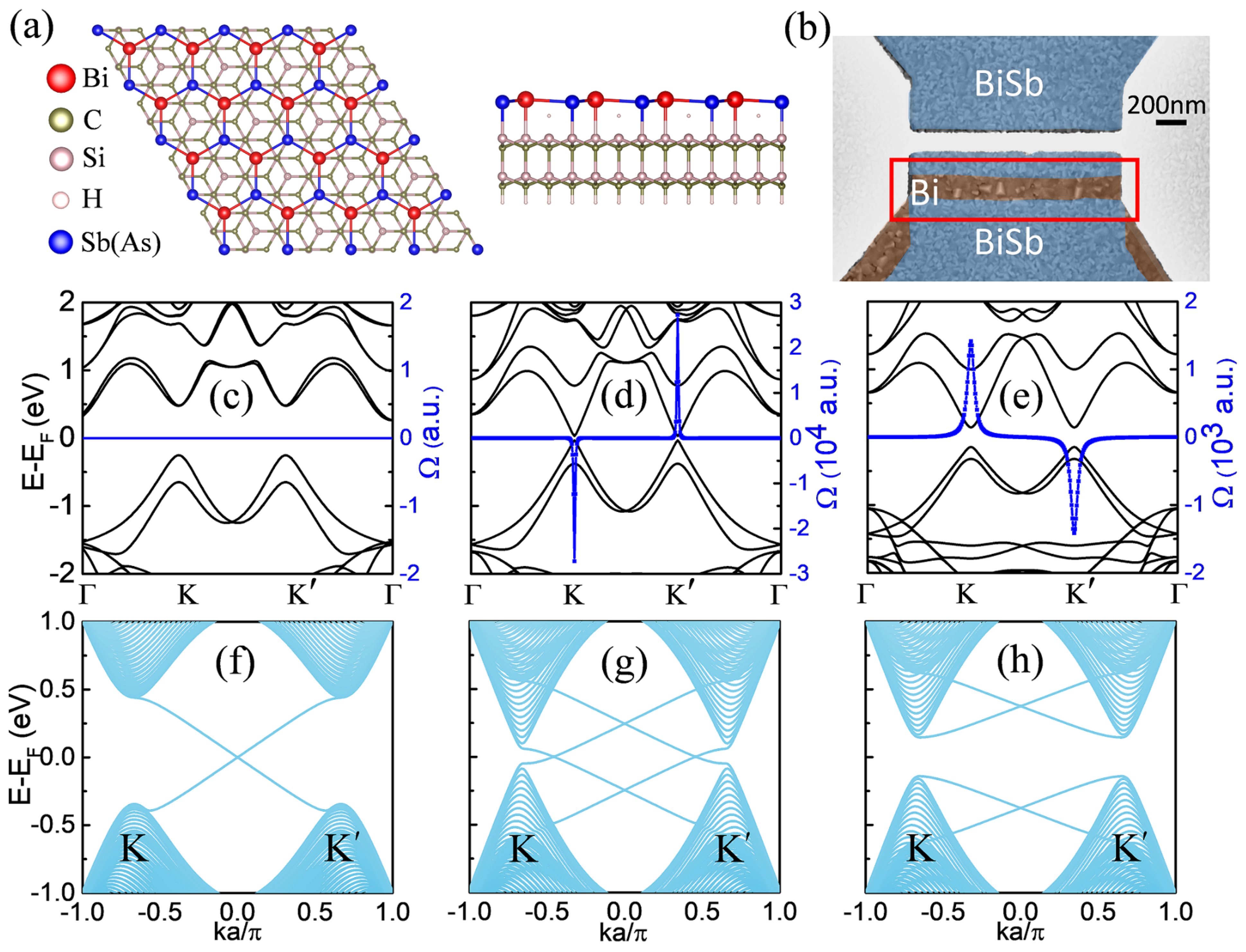}
\caption{(a) Top and side views of a ML BiSb or BiAs on a SiC substrate. (b) Scanning electron micrograph image of the planar BiSb-Bi-BiSb junction with a Si$_3$N$_4$ stencil mask (grey) 300 nm above the film for shadowing. BiSb-Bi interfaces are marked by the red rectangle. (c)-(e) Bands (black) and  Berry curvatures, $\mathbf{\Omega}(\mathbf{k})$ (blue), of the valence bands for the Bi/SiC, BiSb/SiC, and BiAs/SiC, respectively. (f)-(h) Bands of the zigzag nanoribbons for the Bi/SiC, BiSb/SiC, and BiAs/SiC, respectively. The fitted parameters ($\lambda_{SO}$, $U$) for Bi/SiC, BiSb/SiC, and BiAs/SiC are (0.44 eV, 0 eV), (0.30 eV, 0.26 eV), and (0.24 eV, 0.38 eV), respectively.} 
\label{fig:F3}
\end{figure}

{\em Material design}.
The key factor to achieve QSVHK is creating an interface of the QSH and QVH. Since there are large number of hexagonal QSH insulators~\cite{Ren2016:RPP}, a natural way to obtain such an interface is to engineer a part of QSH insulator into a QVH region, where $U > \lambda_{SO}$ is required. Recently, group-V MLs bismuthene, antimonene, and arsenene on a SiC substrate were predicted to be high-temperature 2D topological 
insulators~\cite{Li2018:PRB}. 
For Bi/SiC, a huge nontrivial gap of 0.8$\,$eV has been measured~\cite{Reis2017:S}, originating from the intrinsic SOC of Bi $p_{x,y}$ orbitals~\cite{SMkink}. However, with its inversion symmetry, Bi/SiC fails to show valley-dependent effects, as verified by  
$\mathbf{\Omega}(\mathbf{k})=\mathbf{0}$ at all $\mathbf{k}$ [Fig.~3(c)]. To break the inversion symmetry, we propose to use alloy engineering to induce $U$ in bismuthene, a well-established approach to tailor electronic and topological properties~\cite{Hsieh2008:N,Huang2018:PRL}. Specifically, we propose to grow binary group-V MLs BiSb or BiAs on the SiC substrate, depicted in Fig.~3(a). We expect the change in the binary composition alters the strength of SOC (growing with the atomic number Z) and $U$ (growing with a relative difference in Z of the two group-V elements), thus favoring either QVH or QSH insulators, as shown in Fig.~1(a). BiSb and BiAs films can be fabricated using molecular beam epitaxy (MBE) [Fig. 3(b)], similar to that growth of Bi/SiC or exfoliated from bulk~\cite{Zhang2018:CSR,Gui2019:JMCA}. From first-principles calculations, we see the BiSb/SiC and BiAs/SiC bands near $E_F$ can be accurately described by the Hamiltonian in Eq.~(1)~\cite{SMkink}. 

Without considering SOC, Bi/SiC has gapless Dirac bands at two valleys, while the trivial gaps of 0.52$\,$eV and 0.76$\,$eV are opened in BiSb/SiC and BiAs/SiC~\cite{SMkink}, respectively. Such gaps, originating from the staggered potential, give $U_\mathrm{BiSb/SiC}$ = 0.26$\,$eV and $U_\mathrm{BiAs/SiC}$ = 0.38$\,$eV. With SOC, a nontrivial gap of 66$\,$meV is opened in BiSb/SiC with $\mathbf{\Omega}(\mathbf{k})\neq \mathbf{0}$ [Fig.~3(d)], giving a QSH insulator with $Z_2 = 1$ and the helical edge states [Fig.~3(g)]. The edge states outside the gap are not useful for the robust dissipationless transport, because they are negligible compared to the huge contribution from the trivial bulk bands~\cite{Qi2011:RMP}. Figure~3(e) reveals a different situation for BiAs/SiC. Due to $U > \lambda_{SO}$, a gap of 287$\,$meV appears at $K$ and $K'$ with $Z_2=0$ and no topological edge states [Fig.~3(h)]. Compared to BiSb/SiC, the sign reversal of $\mathbf{\Omega}(\mathbf{k})$ for BiAs/SiC gives the desired QVH phase. 

The resulting QSH-QVH junction [Fig.~1] can be realized combining BiAs/SiC (QVH) with Bi/SiC (QSH) or BiSb/SiC (QSH). Alternatively, to simplify the fabrication and yield QSH with an even larger nontrivial gap, BiAs-Bi/SiC junction is desirable, where the verified QSVHK states are shown in Fig.~2(a). In this analysis we exclude Rashba SOC~\cite{Zutic2004:RMP}, since its influence is negligible in QSVHK as discussed in the SM~\cite{SMkink}. The BiAs-Bi/SiC junction provides a robust platform for QSVHK, protected by a global gap of 287$\,$meV, which is $\sim$ 14 times larger than in BLG ~\cite{Li2016:NN}, supporting ballistic transport at high temperatures, verified by the finite-temperature Green-function calculations and dicussion about the influence of the many-body interaction~\cite{SMkink}. 

The desired QSH-QVH junction can be fabricated using our well-established MBE selective area growth and stencil lithography~\cite{Schuffelgen2019:NN} as shown in Fig. 3(b). The fabrication process and the influence of the stoichiometry are demonstrated in~\cite{SMkink}. Multiple QSH-QVH boundaries can be  created by spatially-selective deposition~\cite{Chen2005:AFM,Schuffelgen2019:NN}, enabling transport of high-density channels. QSVHK robust against nonmagnetic and long-range disorder and insensitive to the interface configurations~\cite{SMkink}, facilitates its experimental observation and possible applications. Unlike QVHK in BLG, the QSVHK in bismuthene is spin-polarized and requires no external field. This offers non-volatility in unexplored applications coupling spin and valley, going beyond low-temperature BLG valleytronic applications~\cite{Li2018:S}. For example, QSVHK supports fully spin-polarized quantum valley currents, making spin-valley filters, valves, and  waveguides possible, or extend the fuctionalities for spin interconnects~\cite{Dery2011:APL,Zutic2019:MT,Lindemann2019:N}. 

\begin{figure}[t]
\centering
\includegraphics[width=0.46\textwidth]{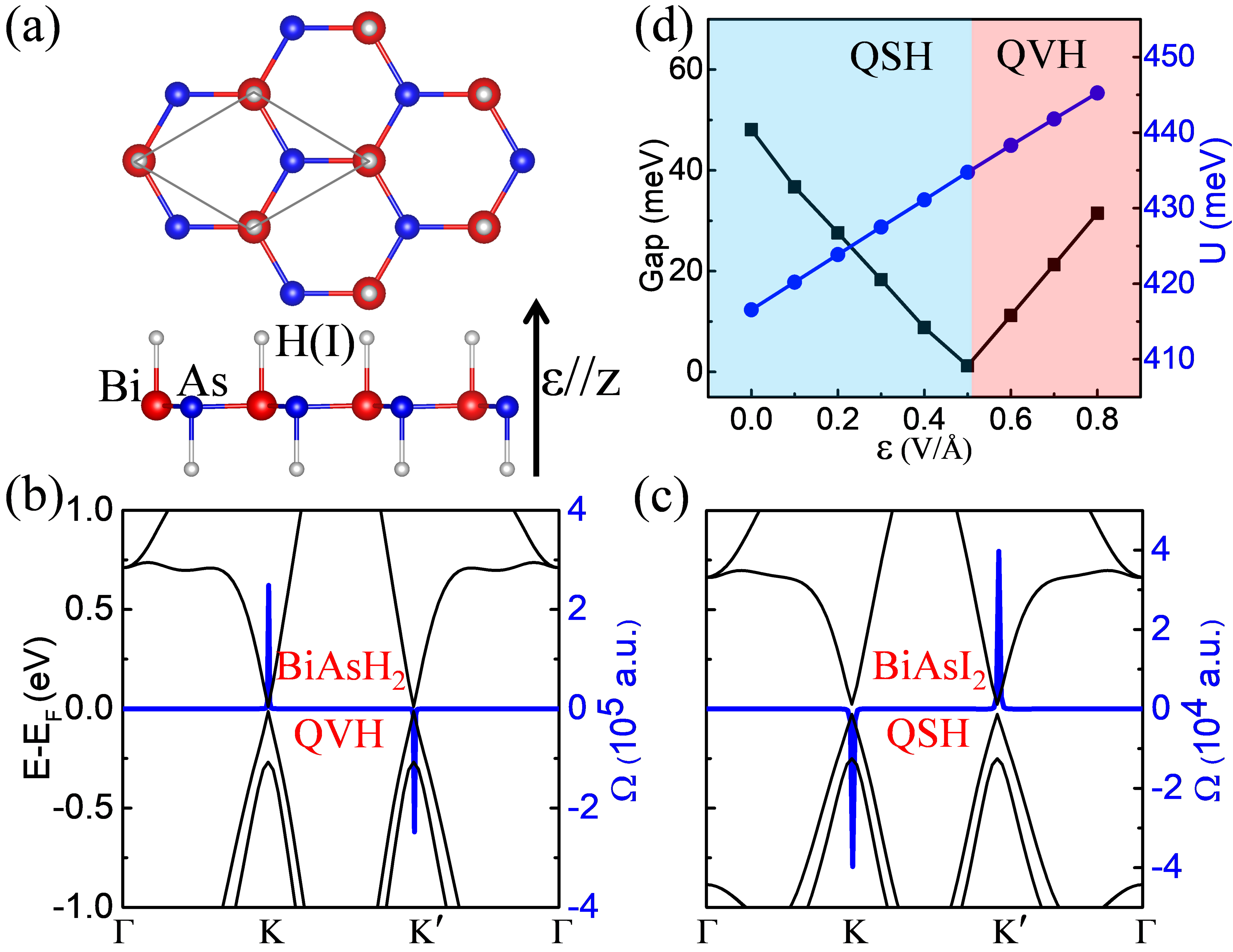}
\caption{(a) Top and side views of the MLs BiAsH$_2$ or BiAsI$_2$. (b) and (c) Bands (black) with Berry curvatures (blue) for MLs BiAsH$_2$ and BiAsI$_2$. (d) Electric-field dependent gap in a ML BiAsI$_2$.}
\label{fig:F4}
\end{figure}
 
Another way to realize QSVHK in bismuthene system is surface decoration, widely used to modify the properties of the 2D materials~\cite{Mannix2017:NRC}. Particularly, hydrogenation and halogenation have been a powerful tool to induce large-gap QSH states in group-IV and V MLs~\cite{Xu2013:PRL,Song2014:NPGAM}. 
Based on first-principles calculations, we show in Fig.~4 that the $\lambda_{SO}$ and $U$ in MLs BiAs can be tuned by the surface decoration, giving either a QSH or a QVH insulator. The structure of the hydrogenated (BiAsH$_2$) and halogenated (BiAsI$_2$) BiAs MLs are shown in Fig.~4(a). From the calculated bands and $\mathbf{\Omega}(\mathbf{k})$ in Figs.~4(b) and (c), we see the desired difference between BiAsH$_2$ and BiAsI$_2$.
While the first is a QVH insulator with a trivial gap of 26$\,$meV, $Z_2=0$, and $\mathbf{\Omega}(\mathbf{k}) \neq \mathbf{0}$ at $K$ and $K'$, the second BiAsI$_2$ is a QSH insulator with a nontrivial gap of 49$\,$meV, 
$Z_2=1$, and reversed $\mathbf{\Omega}(\mathbf{k})$. When such two MLs form a junction [Fig.~1], the QSVHK can emerge along its interface. With hydrogenated and halogenated graphene routinely fabricated~\cite{Balog2010:NM,Karlicky2013:ACSN}, the BiAsH$_2$-BiAsI$_2$ junction could be obtained from ML BiAs to support QSVHK by using the spatially-selective growth and stencil lithography~\cite{SMkink}. 
Since the electric field, $\varepsilon$, can directly change $U$ in 2D materials~\cite{Li2016:NN,Li2018:S}, we also explore the possibility of $\varepsilon$-controlled QSVHK. Figure~4(d) shows that for $\varepsilon$ applied along the z direction in ML BiAsI$_2$, $U$ is increased and the gap is closed when $\varepsilon = 0.5 \:V  / \AA$, the value achievable with ion-liquid gating~\cite{Hebard1987:IEEE,Mannhart1993:APL}. 
Such a gap closing indicates a topological transition from QSH to QVH. Thus, the electric field can also be used to generate and control the QSVHK. 

With experimental realization of the QSVHK it would be possible to verify their inherent robustness of quantized conductance of spin-polarized channels, in contrast to QSH insulators, where this quantization is fragile even at He temperatures~\cite{Konig2007:S,Culcer2020:2DM}.
Furthermore, QSVHK offers an intriguing 
opportunity to study its manifestations of topological superconductivity through proximity effects~\cite{Culcer2020:2DM,Fu2008:PRL,Zutic2019:MT} and test the related role of disorder~\cite{Adagideli2014:PRB,Habibi2018:PRB}.

\begin{acknowledgements}
We thank Prof. Fan Zhang for fruitful discussions.
This work is supported by the U.S. DOE, Office of Science BES, Award No. DE-SC0004890 (T. Z. and I. \v{Z}.), NSFC under Grant Nos. 11874298 (S. C.), 11822407 (H. J.), 11874117 (Z. Y.), and 11574051 (Z. Y.), German Federal Ministry of Education and Research (BMBF) via the Quantum Futur project ``MajoranaChips" (Grant No. 13N15264) within the funding program Photonic Research Germany (M. S. and P. S.) and the UB  Center for Computational Research.
\end{acknowledgements} 
\bibliography{Kink_bib_0812}
\end{document}